%% file: ritik.tex
\documentclass[12pt]{article}

\input preamble.tex

\title{A Natural Language Processing Approach to Malware Classification}

\author{Ritik Mehta\footnotemark[1]\ \ \ 
Olha  Jure\vv{c}kov\'{a}\footnotemark[2]\ \ \ 
Mark Stamp\footnotemark[1]\,\,\footnotemark[3]}

\begin{document}

\symbolfootnotetext[1]{Department of Computer Science, San Jose State University}
\symbolfootnotetext[2]{Faculty of Information Technology, Czech Technical University in Prague}
\symbolfootnotetext[3]{mark.stamp$@$sjsu.edu}

\maketitle

\abstract
Many different machine learning and deep learning techniques
have been successfully employed for malware detection and classification.
Examples of popular learning techniques in the malware domain include
Hidden Markov Models (HMM), Random Forests (RF), Convolutional Neural Networks (CNN), 
Support Vector Machines (SVM), and Recurrent Neural Networks (RNN) 
such as Long Short-Term Memory (LSTM) networks. 
In this research, we consider a hybrid architecture, where HMMs are trained on opcode sequences, 
and the resulting hidden states of these trained HMMs are used as feature vectors in various classifiers.
In this context, extracting the HMM hidden state sequences can be viewed as a form of feature engineering 
that is somewhat analogous to techniques that are commonly employed in Natural Language Processing (NLP). 
We find that this NLP-based approach outperforms other popular techniques on a challenging malware dataset, 
with an HMM-Random Forrest model yielding the best results.

\section{Introduction}

Malware is software that harms or interferes with computer systems. 
Examples of malware include computer viruses, Trojan horses, worms, spyware, ransomware, and adware. 
Despite advances in cybersecurity, malware remains one of the most potent hazards in the cyber environment. 
According to a report published by Sonicwall~\cite{sonicwall}, the number of malware attacks 
worldwide in 2022 was~5.5 billion, a~2\% increase from 2021. 
Better malware detection and classification is required given this escalating trend.

Signature-based techniques are commonly used by anti-virus (AV) applications~\cite{wolpin}. Signature-based detection 
involves security systems creating signatures for patterns observed in malicious software files, so that AV 
applications can efficiently scan for malware. This strategy focuses on individual or small groups of malware samples and it is effective 
against traditional malware. However, there are significant limitations to signature scanning, as it can only cope with 
known malware samples, and numerous code obfuscation techniques have been developed that can 
defeat signature scans: dead code insertion, register reassignment, instruction substitution, and code manipulation 
are some examples of code obfuscation techniques~\cite{obfuscationTechniques}. As a result, 
a large percentage of modern malware can evade signature-based detection. Furthermore, 
extracting signatures for signature-based malware detection requires significant time
and effort~\cite{sigBasedTech2}.

An alternative to signature scanning is heuristic analysis~\cite{heuristicanalysis}. 
However, heuristic analysis has its own set of drawbacks, as heuristics must be carefully 
tweaked to give the best possible identification of emerging threats, while avoiding excessive
false positives on benign code.

Relatively recently, researchers have started using machine learning approaches to detect and analyze malware. 
A wide variety of classic machine learning techniques, including Hidden Markov Models (HMM)~\cite{hmm},
Random Forests (RF)~\cite{randomforest}, and
Support Vector Machines (SVM)~\cite{svm} have been successfully employed in the malware domain. In addition,
deep learning techniques, such as Multilayer Perceptrons (MLP)~\cite{ann} and Long Short-Term Memory (LSTM)~\cite{lstm}
networks have been found to be effective for malware classification. These techniques can be trained on static or dynamic features,
or a combination thereof~\cite{Anusha}. Static features are those that can be obtained without executing or emulating
the code, while dynamic features require execution or emulation. Examples of popular static features are opcode 
sequences and byte~$n$-grams, while an example of dynamic features is API calls. In general, models that 
rely on static features are more efficient as such features are easy to extract and have low computation complexity, 
while models that use dynamic features are more resistant to common obfuscation techniques. 
In this paper, we only consider static features.

In our experiments, we consider hybrid machine learning techniques, where we first train HMMs on opcode sequences,
then we determine the hidden state sequences from the trained HMMs, and, finally, we classify malware samples
into their respective families based on these hidden state sequences. That is, the HMM training serves as a 
feature engineering step
that uncovers ``hidden'' information in the opcode sequence, in an approach that is analogous 
to techniques that are often used in Natural Language Processing (NLP) applications. For example, 
when determining the part of speech of words in English sentences (noun, verb, adjective, adverb, etc.),
we could train an HMM, then use the hidden states to determine the most likely classification for each 
individual word. In our malware experiments, we consider a variety of classifiers and find that a RF
performs best on our derived hidden state sequence---we refer to the resulting hybrid model as 
an HMM-RF.

The remainder of this paper consists of the following. In Section~\ref{sect:back} we 
present relevant background information, a brief introduction
to the learning techniques considered in our research, and a selective survey of some
relevant previous work. Section~\ref{sect:meth} covers our experimental
design and provides a brief description of the dataset used, while Section~\ref{sect:exp} 
gives our experimental results. We conclude the paper 
with Section~\ref{sect:conc}, which includes some ideas for future work.

\section{Background}\label{sect:back}

In this section, we first introduce the learning techniques that appear in subsequent sections of this paper.
Then we provide a brief overview of a few of the most relevant related research papers.

\subsection{Hidden Markov Model (HMM)}

\begin{figure}[!htb]
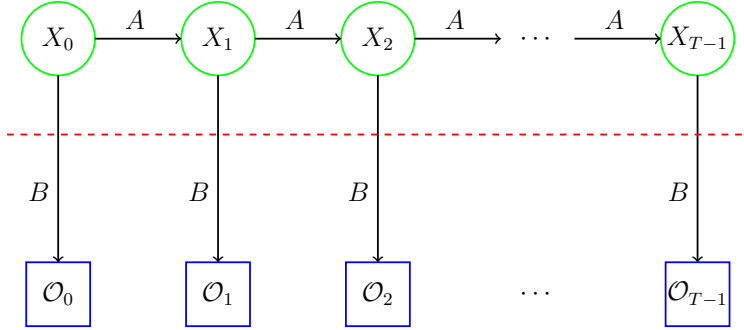

\centering
\adjustbox{scale=0.85}{
\input figures/hmm.tex
}
\caption{Hidden Markov Model~\cite{introStamp}}\label{fig:HMM}
\end{figure}

Hidden Markov Models (HMM)~\cite{hmm} can be described as statistical Markov models 
in which the states are hidden. An HMM can be represented as $\lambda=(A, B, \pi)$, 
where~$A$ is the state transition probability matrix, $B$ is the observation probability matrix, 
and~$\pi$ is the initial state distribution. 
A series of observations, denoted as~$\cO$, are available, and these observations are
probabilistically related to the hidden states sequence~$X$ via the~$B$ matrix.
Figure~\ref{fig:HMM} provides a high-level view of an HMM.

The number of hidden states in an HMM is denoted as~$N$ and the number of unique observation symbols is
denoted as~$M$, while the length of the observation sequence is~$T$. Thus, the~$A$ matrix
is~$N\times N$, the~$B$ matrix is~$N\times M$, and the observation and hidden state sequences
are of length~$T$. Furthermore, each row of the~$A$, $B$, and~$\pi$ matrices 
is row-stochastic, that is, each row represents a discrete probability distribution.

Given a hidden state sequence~$X = (x_0, x_1, \ldots, x_{T-1})$, 
and the corresponding observation sequence~$\cO = (\cO_0, \cO_1, \ldots, \cO_{T-1})$, 
the probability of state sequence~$X$ is given by
\begin{align*}
  P(X, \cO) =& \pi_{x_0}b_{x_0}(\cO_0)a_{x_0, x_1}b_{x_1}(\cO_1)a_{x_1, x_2}\cdots \\
      & \ \ \ \ \ \ \ \ \ \ \cdots b_{x_{T-2}}(\cO_{T-2})a_{x_{T-2}, x_{T-1}}b_{x_{T-1}}(\cO_{T-1})
\end{align*}
where~$a_{x_i, x_j}$ is the state transition probability from~$x_i$ to~$x_j$, 
$b_{x_i}(\cO_i)$ is the probability of observing~$\cO_i$ in the hidden state~$x_i$, 
and~$\pi_{x_0}$ is the probability of starting in state~$x_0$. 

Within the HMM framework, there are efficient algorithms to solve the following three problems~\cite{mshmm}.
\begin{enumerate}
\item Given an HMM~$\lambda = (A, B, \pi)$ and an observation sequence~$\cO$, 
we can compute a score of the observation sequence~$\cO$ with respect to the model~$\lambda$,
where the score is based on the conditional probability~$P(\cO\,|\,\lambda)$.
\item Given a model~$\lambda = (A, B, \pi)$ and an observation sequence~$\cO$, we can determine 
the optimal hidden state sequence corresponding to~$\cO$, where ``optimal'' is defined as maximizing
the expected number of correct states. Note that this implies HMMs are an Expectation Maximization (EM)
technique, and that the HMM solution to this problem differs, in general, from a dynamic program, where
we maximize with respect to the overall path.
\item Given an observation sequence~$\cO$ and a specified number of hidden states~$N$, we can
train an HMM. That is, we can determine the matrices that comprise the model~$\lambda = (A, B, \pi)$,
so that~$P(\cO\,|\,\lambda)$ is maximized. 
\end{enumerate}
The efficient solution to problem~1 relies on the so-called forward algorithm, while the forward algorithm and the
backward algorithm enable an efficient meet-in-the-middle approach to solve problem~2~\cite{hmmForwardAndBackward}. 
Typically, the Baum-Welch re-estimation algorithm, which is a hill climb technique,
is used to train an HMM to model a given observation sequence. 
For the research in this paper, we will be focusing on the solutions to problem~2 and~3.

\subsection{Random Forest}

\begin{figure}[!htb]
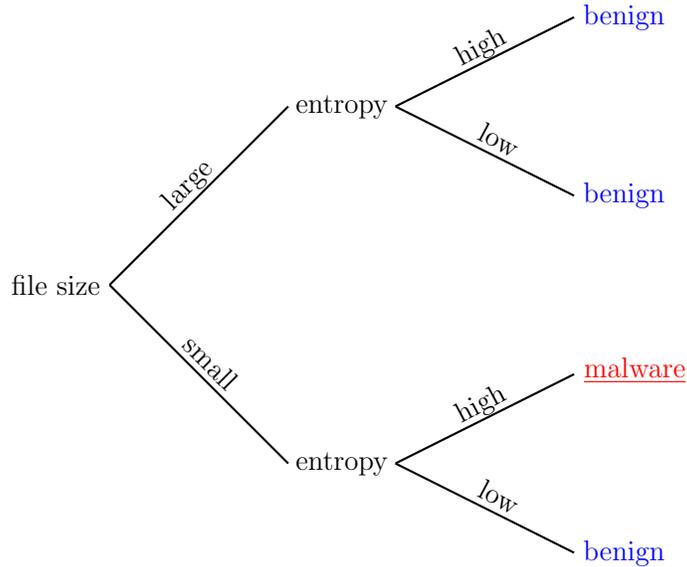

\centering
\input figures/tree.tex
\caption{A decision tree example~\cite{introStamp}}\label{fig:decisionTree}
\end{figure}

Random Forests (RF), which were originally proposed in~\cite{randomforestorig} in~2001,
consists of ensembles of decision trees.
A decision tree~\cite{decisionTree} is a simple supervised technique that is employed to categorize 
or make predictions based on a tree structure. A decision tree is comprised of a root node, branches, 
internal nodes, and leaf nodes. Decision trees can be regarded as a collection of \texttt{if}-\texttt{else} 
statements. For example, Figure~\ref{fig:decisionTree} represents a decision tree to determine whether 
a sample is malware or not, based on two features, namely, file size and entropy. In this example, 
files that are small and have high entropy are classified as malware, where the thresholds for
``small'' size and ``high'' entropy would be determined based on training data.

In a decision tree, we want to make more important or informative decisions closer to the root node,
as this makes the most efficient use of the information available, and it allows for
pruning of trees with minimal loss of accuracy. Information Gain (IG) is a popular
measure of the importance of a feature. IG is defined as
$$
  \mbox{IG} = H(S) - H(S\,|\,X)
$$
where~$H$ represents entropy, which is a measure of uncertainty. 
Here, $H(S)$ is the entropy of the original dataset~$S$, while~$H(S\,|\,X)$ represents 
the conditional entropy of~$S$, given the value of a specific feature~$X$.
Entropy, in turn, is defined as
$$
  H(S) = \sum_{i=1}^{n} -p_{i}\log p_{i}
$$
where~$n$ is the number of classes and~$p_i$ is the probability of 
a sample belonging to class~$i$. 
Note that instead of IG, other criteria, such as
the Gini coefficient~\cite{gini}, are sometimes used
when constructing decision trees.

As mentioned above, a Random Forest~\cite{randomforest} (RF) is an ensemble machine learning technique 
that is based on a collection of decision trees. RFs can be used for both classification and regression tasks. 
The first step in the RF algorithm consists of selecting a subset of features and data points for constructing each 
decision tree. Each decision tree will produce an output, with the final result of the RF is based on a majority vote 
or averaging scheme for classification or regression, respectively. 

In this research, we consider following three important hyperparameters of an RF. 
\begin{itemize}
    \item The number of decision trees in the RF, which is denoted as \texttt{n\_estimators}.
    \item The maximum number of features to considered while looking for the best split,
    denoted as \texttt{max\_features}.
    \item The function that evaluates the quality of a specific split, denoted as the \texttt{criterion}.
    Specifically, we consider Gini and entropy.
\end{itemize}

\subsection{Literature Review}\label{sect:review}

There has been a substantial amount of previous work on malware classification using 
a wide range of machine learning approaches. 
This section discusses a representative sampling 
of such malware classifications techniques, with the emphasis on
research that is most similar to our novel NLP-based HMM-RF technique.

\subsection{Malware Classification using HMM}

In one of the earliest papers in this genre, Wong and Stamp~\cite{Wing} consider HMMs for the detection
of metamorphic malware. By modern standards, they considered a very small sample set, but they
were able to distinguish malware from benign with high accuracy, clearly indicating the viability of
machine learning models within the malware domain.

Annachhatre et al.~\cite{hmmmalwareclass} train multiple HMMs on a variety of metamorphic malware samples.
Each malware sample in the test set is then scored against all models, and the samples are clustered
based on the resulting vector of scores.
They were able to classify the malware samples into their respective families with good accuracy,
based solely on the clustering results, and they even obtained good accuracy on malware samples 
belonging to families for which no explicit model had been trained.

In~\cite{gmmhmm}, Zhao et al., explore the usage of complex Gaussian Mixture Model-HMMs (GMM-HMM) 
for malware classification. In their research, GMM-HMMs produced comparable results to discrete HMMs 
based on opcode sequence features, and showed significant improvement over discrete HMMs when trained on 
entropy-based features.

\subsection{Malware Classification using SVM}

Support Vector Machines (SVM) are a prominent class of techniques for supervised learning. 
The objective of the SVM algorithm is to determine an optimal hyperplane---or hyperplanes, in the
the more general multiclass case---that can segregate $n$-dimensional space 
into classes. The decision boundary is then used to classify data points not in the training set. 
In~\cite{svm1}, Kruczkowski et al., trained an SVM on malware samples and achieved a 
cross-validation accuracy of~0.9398, and an F1-score of~0.9552. 

Singh et al.~\cite{svm2} also use SVMs for malware classification. They trained 
HMMs, computed a Simple Substitution Distance (SSD) score based on the classic encryption 
technique from symmetric cryptography, and also computed an Opcode Graph Score (OGS). 
Each malware sample was classified---using an SVM---based on its vector of these three scores. 
While the individual scores generally performing poorly in a robustness analysis, 
the SVM results were significantly more robust, indicating the advantage of combining 
multiple scores via an SVM.

\subsection{Malware Classification using Random Forest}

In~\cite{randomforest1}, Garcia and Muga~II employ an approach for converting a binary file to a gray scale image, 
and subsequently use an RF to classify malware into families, with an accuracy of~0.9562 being achieved.
Domenick et al.~\cite{randomforest2}, on the other hand, combine an RF with 
Principal Component Analysis (PCA)~\cite{pca} and Term Frequency-Inverse Document Frequency (TF-IDF)~\cite{tfidf}.
The model based on RF and PCA outperformed a models based on Logistic Regression, 
Decision Trees, and SVM on one datasets, while the model based on Random Forest and TF-IDF performed 
best on a second dataset.

\subsection{Malware Classification using RNN and LSTM}

A Recurrent Neural Network (RNN)~\cite{rnnRef} is a type of neural network designed 
to process sequential data by incorporating feedback connections. 
However, generic RNNs are subject to computational issues, including
vanishing and exploding gradients, which limit their utility. Consequently, various specialized RNN-based
architectures have been developed, which mitigate some of the issues observed in plain vanilla RNNs.
The best-known and most successful of these specialized RNN architectures is the 
Long Short-Term Memory (LSTM) model.

An unsupervised approach of using Echo State Networks (ESNs)~\cite{esns} and RNNs for a 
``projection'' stage to extract features is discussed by Pascanu et al., in~\cite{malwareclassrnn}. 
A standard classifier then uses these extracted features to detect malicious samples. Their hybrid model 
with the best performance employed ESN for the recurrent model, a max pooling layer 
for non-linear sampling, and Logistic Regression for the final classification.

R. Lu, in~\cite{lstmmalwareclass}, experimented with LSTMs for malware classification. 
First, Word2Vec word embedding of the opcodes were generated using skip-gram and CBOW models. 
Subsequently, a two stage LSTM model was used for malware detection. The two-stage LSTM model 
is composed of two LSTM layers and one mean-pooling layer to obtain feature representations of 
malware opcode sequences. An average AUC of~0.987 was achieved for malware 
classification on a modest-sized dataset consisting of~969 malware and~123 benign files.

\subsection{Malware Classification using CNN}

Recently, image-based analysis of malware has been the focus of considerable
research; see~\cite{Niket,Mugdha,Huy,Sravani}, for examples. Much of the work is based
on Convolutional Neural Networks (CNN)~\cite{cnn}. A CNN is a type of neural network that designed
to efficiently deal with data that is in a grid-like layout where local structure dominates, such as 
is the case in images. In~\cite{cnn-mal-class}, Kalash et al., proposed a CNN-based architecture, 
called M-CNN, for malware classification. The architecture of M-CNN is based on the 
VGG-16~\cite{vgg16}, and it achieves accuracies of~0.9852 and~0.9997 on the 
popular MalImg~\cite{malimg} dataset and a Microsoft~\cite{microsoft-dataset} dataset, 
respectively.

\section{Methodology}\label{sect:meth}

In this section, we first introduce the dataset used in our experiments. We then outline
the experimental design that we employ for the experiments presented in Section~\ref{sect:exp}.

\subsection{Dataset and Preprocessing}

\begin{figure}[!htb]
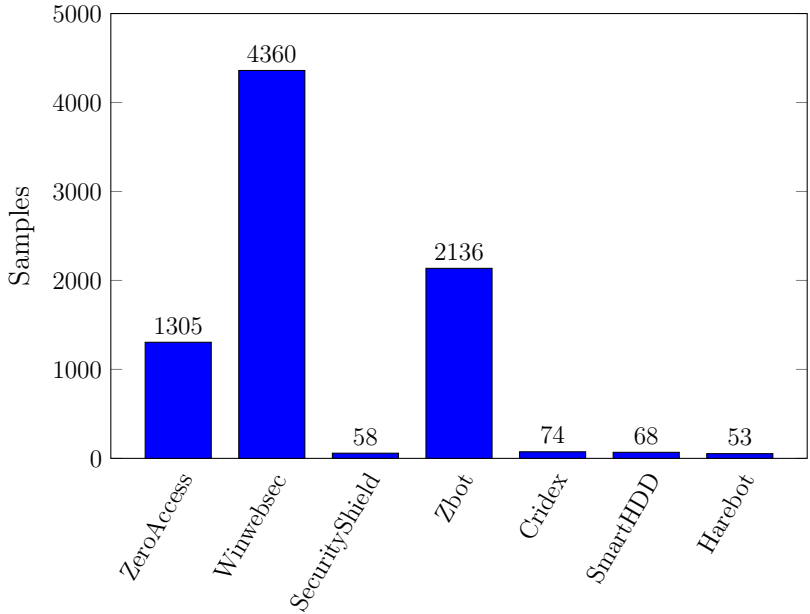

\centering
\input figures/dist.tex
\caption{Malware samples per family}\label{fig:Dataset distribution}
\end{figure}

In this research, we use the well-known Malicia dataset~\cite{Malicia}. 
The dataset includes~48 different malware families. However, the dataset is highly imbalanced,
and we removed all classes with less than~50 samples. This results in malware 
samples belonging to~7 classes. For our experiments, we use an 80-20 train-test split, 
i.e., 80\% of the samples are used for training, while 20\% of the samples are used for testing. 
The distribution of samples in the malware families is 
shown in Figure~\ref{fig:Dataset distribution}.

\subsection{Experimental Design}

The methodology for training our HMM-RF model 
can be summarized as follows. 
\begin{enumerate}
    \item Train HMMs on opcode sequences --- This step consists of training~$n$ different HMMs, 
    where~$n$ is the number of classes. Each HMM is trained based on the opcode sequences of 
    samples belonging to a particular class. Recall that when training an HMM, 
    we specify the number of hidden states~$N$.
    \item Determine the hidden state sequence for each sample --- The first~$L$ opcodes 
    of a given sample are fed into each HMM. 
    This results in $n$ hidden state sequence vectors that are of length~$L$.
    \item Concatenating the hidden state sequences --- For each malware sample, we concatenate the $n$ hidden state sequences obtained in the previous step.
    \item Scale the hidden state sequences --- In this step, each concatenated hidden state sequence vector is scaled using a standard scaler.
    \item Train the RF model --- We then train an RF model using the scaled hidden state sequences obtained in the previous step
    as the feature vectors. Of course, the malware family to which the sample belongs serves as the corresponding label.
\end{enumerate}
To summarize, we train an HMM for each family, then use the trained HMMs to determine the
hidden state sequences corresponding to each sample, with these hidden state sequences then used
to train an RF model. For any new malware sample, we first generate the concatenated hidden state 
sequence by feeding the first~$L$ opcodes to each HMM. The next step is to scale the concatenated 
hidden state sequence, and subsequently use the Random Forest model to determine the class of the malware sample. 
The motivation for this approach comes from Natural Language Processing (NLP),
where uncovering the hidden state sequence is a fundamental step in analyzing text. As far as the 
authors are aware, this NLP-based approach has not previously been employed in the malware domain.

\section{Experiments and Results}\label{sect:exp}

In this section, we first discuss the training of the HMMs and their use to obtain
hidden state sequences, and we consider the training of our HMM-RF classifier,
including hyperparameter tuning.
We then summarize the results of our experiments, and we compare these
results to other other similar models. We conclude this section with a comparison of 
our results to other research involving the Malicia dataset.

\subsection{HMM Training and Hidden States}

As discussed above, the subset of the Malicia dataset that we use
consists of seven malware families. We train one HMM for each family,
and hence we have seven trained HMMs, where each model is of the form~$\lambda=(A, B, \pi)$.
We experimented with~$N\in\{5,10,20,30\}$, where~$N$ is the number of hidden states, and we found that~20
yields the best results. The number of unique observations (i.e., a superset of the opcodes in all seven families)
is~$426$, with \textit{MOV} being the most frequent. Hence, $N=20$ and~$M = 426$ in all of our models. 

Recall that the HMM matrices are~$A=\{a_{ij}\}$, which is~$N\times N$,
$B=b_i(j)$, which is~$N\times M$, and~$\pi=\{\pi_i\}$, which is~$1\times N$.
We initialize the~$A$, $B$, and~$\pi$ matrices to approximately uniform, that is, 
each~$a_{ij} \approx 1/N$, each~$b_i(j) \approx 1/M$, and each~$\pi_i \approx 1/N$,
while enforcing the row stochastic conditions.
The minimum number of iterations of the Baum-Welch re-estimation algorithm
is set to~10, and we stop when successive iterations beyond this number produce a change in the model
score of less than~$\varepsilon = 0.001$. 
When training our models, the average number of iterations was~10.43, 
and it took an average of five hours to train each HMM.

Next, we use the trained HMMs to generate hidden state sequences for each sample. 
Given a sample, we generate a hidden state sequence using the HMM 
corresponding to the family that the sample belongs to. The length of the hidden state sequence 
corresponding to each malware sample is truncated to a constant~$L$, that is, we only use the 
hidden states corresponding to the first~$L$ opcodes. We experiment with~$L\in\{25,50,100,200\}$.
In rare cases there were insufficient opcodes available in a given sample, 
i.e., the length of opcode sequence for the malware sample was less than $L$, 
in which case we dropped
the sample from consideration; the number of such exceptional cases for
each value of~$L$ is given in Table~\ref{tab: malware_sample_dropped}.
As can be observed from Figure~\ref{fig:Dataset distribution}, the total number 
of malware samples in the seven classed is~8054, and hence an insignificant
percentage of malware samples were dropped for each value of~$L$.

\begin{table}[!htb]
\centering
\caption{Number of malware sample dropped for different values of $L$}
\label{tab: malware_sample_dropped}
\adjustbox{scale=0.85}{%
\begin{tabular}{c|c}\midrule\midrule
$L$ & Samples dropped\\ \midrule
25 & 3 \\
50 & 11 \\
100 & 14 \\
200 & 26 \\ \midrule\midrule
\end{tabular}
}
\end{table}

\subsection{HMM-RF Training}

As discussed above, in our HMM-RF, a standard Random Forests algorithm is trained 
on the hidden state sequences generated by HMMs. For each sample, we generate the 
hidden state sequence of length~$L$ for each of the seven trained HMMs. These
hidden state sequences are then concatenated, yielding a feature vector of length~$7L$
for each sample.

Using the feature vectors discussed in the previous paragraph, we 
conducted a grid search~\cite{grid-search} to determine the hyperparameters of 
our RF classifier. We tested the hyperparameter values in 
Table~\ref{tab: param_HMMRF}, 
with the values in boldface yielding the best result.

\begin{table}[!htb]
\centering
\caption{HMM-RF hyperparameters tested}\label{tab: param_HMMRF}
\adjustbox{scale=0.85}{
\begin{tabular}{c|c}\midrule\midrule
Hyperparameter & Values\\ \midrule
$L$ & 25, \textbf{50}, 100, 200\\
\texttt{n\_estimators} & 1, 10, 100, \textbf{150}, 200\\
\texttt{criterion} & \textbf{gini}, entropy, log\_loss\\
\texttt{max\_features} & \textbf{sqrt}, log2, None \\ \midrule\midrule
\end{tabular}
}
\end{table}

\subsubsection{Results}

The accuracy we obtained for the best choice of hyperparameters in Table~\ref{tab: param_HMMRF}
was~0.9758. In Figures~\ref{fig:hmmrfGraph_Features}(a), \ref{fig:hmmrfGraph_Features}(b), 
and~\ref{fig:hmmrfGraph_Features}(c), we give expanded results for each of the individual hyperparameters
in Table~\ref{tab: param_HMMRF}, namely, \texttt{n\_estimators}, \texttt{criterion}, 
and \texttt{max\_features}, respectively. In Figure~\ref{fig:hmmrfGraph_Features}, the hyperparameter to be tested is kept fixed, and an average of the accuracy obtained after running a grid search on other hyperparameters is plotted. For example, in Figure~\ref{fig:hmmrfGraph_Features}(a), \texttt{n\_estimators} is fixed to a particular value, and an average accuracy obtained by running a grid search on \texttt{criterion} and \texttt{max\_features} is plotted to correspond to that value.
It can be observed that, for example, an HMM-RF with~$L = 50$ 
performs better than its counterparts for all other combinations of hyperparameters.


\begin{figure}[!htb]
\centering
\begin{tabular}{cc}
\multicolumn{2}{c}{\input figures/line.tex} 
\\
\multicolumn{2}{c}{\adjustbox{scale=0.9}{(a) \texttt{n\_estimators}}}
\\
\\
\input figures/criterion.tex
&
\input figures/max.tex
\\
(b) \adjustbox{scale=0.9}{\texttt{criterion}}
&
(c) \adjustbox{scale=0.9}{\texttt{max\_features}}
\end{tabular}
\caption{Training accuracy trends for different hyperparameters}\label{fig:hmmrfGraph_Features}
\end{figure}

%
%

\begin{figure}[!htb]
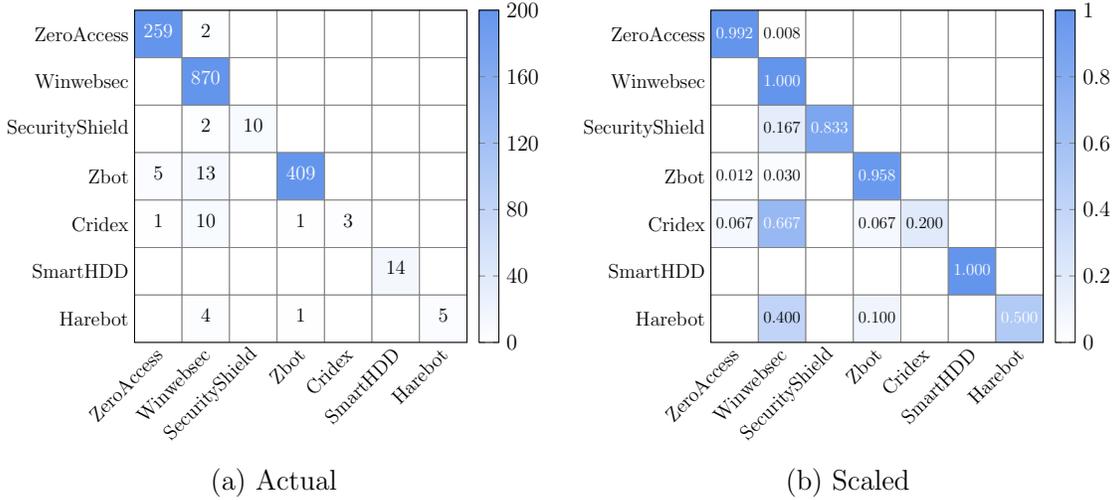

\centering
\begin{tabular}{cc}
\input figures/conf_HMM-RF.tex
&
\input figures/conf_relative_HMM-RF.tex
\\
\adjustbox{scale=0.9}{(a) Actual}
&
\adjustbox{scale=0.9}{(b) Scaled}
\end{tabular}
\caption{Confusion matrice for HMM-RF model}\label{fig:conf_HMM-RF}
\end{figure}

In Figure~\ref{fig:conf_HMM-RF} we provide a pair of confusion matrices for our HMM-RF 
experimental results.
Figure~\ref{fig:conf_HMM-RF}(a) gives the actual number of classifications for each case.
Since the number of samples per class is highly imbalanced, in Figure~\ref{fig:conf_HMM-RF}(b)
we provide a confusion matrix that is scaled to the number of samples per class.
We observe that the samples from the three largest classes, namely, ZeroAccess,
Winwebsec, and Zbot, are classified with an accuracy of 0.9872.
Of the four smaller classes, Cridex and Harebot are poorly classified; however,
the numbers in those classes are so small that they have minimal effect on the overall
classification accuracy.

\subsection{Comparison to Related Techniques}

We compared the results obtained from our HMM-RF model with some related techniques. 
Table~\ref{tab: acc_table} shows the accuracy and weighted 
F1-score obtained after testing the following techniques
on the same seven families of the Malicia dataset. 
\begin{itemize}
\item Word2Vec-LSTM --- For this model, we generated Word2Vec embeddings of the opcodes, and then
trained and LSTM model on the resulting sequence of embedding vectors.
\item BERT-LSTM --- This is the same as the Word2Vec-LSTM model, except that BERT was used to generate
the embedding vectors, instead of Word2Vec.
\item Random Forest --- For this model, we trained a Random Forest model directly on the opcode sequences. 
We obtained the feature vectors by truncating the resulting number sequence to a length $L$. 
Table~\ref{tab:hyperparams} shows the hyperparameters tested for Random Forest model,
with those selected in boldface.
\item SVM --- As with the previous model, this model was also trained on the feature vectors obtained directly 
from the opcode sequences, but we used 
an SVM classifier, instead of a Random Forest. The hyperparameters tested and selected for SVM are shown in 
Table~\ref{tab:hyperparams}, where $L$ is the length of the feature vectors 
and $C$ is the regularization parameter. 
\item HMM-SVM --- This model is the most similar to our HMM-RF model, with the only difference 
being that we trained an SVM as the classifier, rather than using a Random Forest. 
We tested the HMM-SVM model with the same lengths of hidden state sequences as HMM-RF. 
Table~\ref{tab:hyperparams} shows the tested and selected hyperparameters for our HMM-SVM model.
\end{itemize}
More information on the training of the LSTM models can be found in the Appendix~\ref{appendix}. 

%
%

\begin{table}[!htb]
\centering
\caption{Hyperparameters tested and selected}\label{tab:hyperparams}
\adjustbox{scale=0.85}{
\begin{tabular}{c|c|c}\midrule\midrule
Technique & Hyperparameter & Values\\ \midrule
\multirow{4}{*}{RF} & 
$L$ & 25, \textbf{50}, 100, 200\\
& \texttt{n\_estimators} & 1, 10, \textbf{100}, 150\\
& \texttt{criterion} & \textbf{gini}, entropy, log\_loss\\
& \texttt{max\_features} & \textbf{sqrt}, log2, None \\ \midrule
%
%
\multirow{4}{*}{SVM} & 
$L$ & 25, \textbf{50}, 100, 200\\
& $C$ & 0.5, 1, 5, \textbf{10}\\
& \texttt{degree} & \textbf{2}, 3, 4, 5\\
& \texttt{kernel} & linear, poly, \textbf{rbf}, sigmoid \\ \midrule
%
%
\multirow{4}{*}{HMM-SVM} & 
$L$ & 25, \textbf{50}, 100, 200\\
& $C$ & 0.5, 1, \textbf{5}, 10\\
& \texttt{degree} & \textbf{2}, 3, 4, 5\\
& \texttt{kernel} & linear, poly, \textbf{rbf}, sigmoid \\ \midrule\midrule
\end{tabular}
}
\end{table}

We observe that our HMM-RF slightly outperforms the HMM-SVM, with Word2Vec-LSTM, Random Forest, and SVM
models also performing reasonably well. Only the BERT-LSTM embedding does poorly, which is perhaps at least
partially due to insufficient training data for the more complex BERT embedding technique.

\begin{table}[!htb]
\centering
\caption{Classification metrics of different techniques}\label{tab: acc_table}
\adjustbox{scale=0.85}{
\begin{tabular}{c|c|c}\midrule\midrule
\multirow{2}{*}{Technique} & \multicolumn{2}{c}{Validation} \\ \cline{2-3} \\[-2.25ex]
                  & Accuracy & F1-score\\ \midrule
Word2Vec-LSTM & 0.9714 & 0.9658\\
BERT-LSTM & 0.9181 & 0.9037\\
Random Forest & 0.9702 & 0.9668 \\
\textbf{HMM-RF} & \textbf{0.9758} & \textbf{0.9732}\\
SVM & 0.9589 & 0.9535 \\
HMM-SVM & 0.9757 & 0.9727 \\ 
\midrule\midrule
\end{tabular}
}
\end{table}

\subsection{Comparison to Previous Work}

In previous work, many experiments have been performed on the Malicia~\cite{Malicia} dataset. 
Here, we compare our results with previous work done on this same dataset. In~\cite{Sravani} 
malware scores were computed based on image processing. The data was classified into five categories, 
four of which consisted of malware and the last containing benign samples. The authors obtained 
an accuracy of~0.9285 with the 80-20 train-test split. 

The research in~\cite{previouswork2} focuses on accuracy as a function of the size
of the training dataset; in the best cases they obtain 
accuracies of~0.9718 using~$k$-Nearest Neighbor ($k$-NN) 
and~0.9726 using an Artificial Neural Network (ANN). 
In~\cite{gmmhmm} a Gaussian Mixture Model-Hidden Markov Model (GMM-HMM) 
approach for malware classification was found to yield an accuracy of~0.9467. 
The research presented in~\cite{Niket} achieves an accuracy of~0.9293 using transfer learning 
with image-based techniques, based on the five most populous 
classes in the Malicia dataset. We note in passing that in~\cite{Niket}, higher accuracies
are obtained using~$k$-NN, but these results show clear signs of overfitting.

The research in~\cite{previouswork5} includes a large
number of experiments involving Long Short-Term Memory (LSTM) models and
variants thereof. In this case, the models are trained and tested on~20
malware families, but only three of these families are from the Malicia 
dataset. 

Table~\ref{tab: comp_table} summarizes the examples of previous work
discussed in this section. Although some of the accuracies in Table~\ref{tab: comp_table}
are comparable to the accuracy that we obtain, those results are 
for problems that are inherently easier, due to the number of classes
considered.

\begin{table}[!htb]
\centering
\caption{Comparison to previous work}\label{tab: comp_table}
\adjustbox{scale=0.85}{
\begin{tabular}{c|ccc}\midrule\midrule
Research & Technique & Classes & Accuracy\\ \midrule
Bhodia, et. al~\cite{Niket} & Transfer Learning & \zz2 & 0.9761 \\
Bhodia, et. al~\cite{Niket} & Transfer Learning & \zz5 & 0.9293 \\
Dang, et. al~\cite{previouswork5} & MLP & 20 & 0.6069 \\
Dang, et. al~\cite{previouswork5} & LSTM without embedding & 20 & 0.4001 \\
Dang, et. al~\cite{previouswork5} & LSTM with embedding & 20 & 0.5814 \\
Dang, et. al~\cite{previouswork5} & biLSTM & 20 & 0.7946 \\
Dang, et. al~\cite{previouswork5} & biLSTM + embedding + CNN & 20 & 0.8742 \\ 
Jain~\cite{previouswork2} & $k$-NN & \zz3 & 0.9718 \\
Jain~\cite{previouswork2} & ANN & \zz3 & 0.9726 \\
Yajamanam et. al~\cite{Sravani} & Image processing & \zz3 & 0.9300 \\
Zhao, et. al~\cite{gmmhmm} & GMM-HMM & \zz3 & 0.9467 \\ \midrule
Our research & HMM-RF & \zz5 & 0.9758 \\
\midrule\midrule
\end{tabular}
}
\end{table}

\section{Conclusion and Future Work}\label{sect:conc}

In this paper, we focused our attention on a hybrid Hidden Markov Model-Random Forest (HMM-RF)
model. In this model, HMMs were trained on
opcode sequences derived from each of the seven
malware families in our dataset. These models were then used to determine the hidden state sequences for each
sample, and the resulting HMM hidden state sequence vectors were then used as feature vectors in a Random Forest classifier.
We found that our HMM-RF model outperformed several comparable techniques on the same dataset, although an analogous
HMM-SVM technique performed virtually the same, with respect to accuracy. In contrast, techniques that did not use
the HMM hidden state sequences as features performed measurably worse. This indicates that training an HMM
and using it to uncover the hidden states is valuable feature engineering step. The hidden state sequence of HMMs
are often used in NLP applications but, as far as the authors are aware, this approach
has not previously been applied to malware-related problems. Our results indicate that this NLP-based 
technique holds promise in the malware domain, and it would be worth investigating in other domains as well.

There are many possible avenues for future work. Testing on larger and more challenging datasets is always useful.
Testing additional sequential learning techniques on derived hidden state sequences is another area that deserves
further investigation. Recently, image-based analysis of malware has been shown to be highly effective. Applying 
Convolutional Neural Networks to images derived from hidden state sequences might provide a means 
of retaining the apparent feature engineering advantage that we observed in this paper, while also providing
the improved classification results that have been observed using advanced image-based learning models.

\bibliographystyle{plain}
\bibliography{references.bib}

\appendix
\section*{Appendix}\label{appendix}


In this Appendix, we discuss our two LSTM  models in more detail.
The first of these models relies on Word2Vec embeddings, while the second
uses BERT embeddings. We chose Word2Vec and BERT embeddings for our experiments to leverage the benefits of transfer learning and semantic representation.

Table~\ref{tab: summary_lstm_models} depicts the summary of our LSTM models. We trained our LSTM models on fixed-length opcode sequence. Note that an LSTM can be trained on a variable length input sequence. However, a fixed length was chosen for efficient parallel processing. We experimented with the
length parameter, which we denote as~$k$.
For opcode sequence of length greater than~$k$, we simply truncate,
while in the rare cases where the opcode sequence is of length less than~$k$, we
pad the sequence with zeros. Note that our ``vocabulary'' is of size~426,
that is, there are~426 unique opcodes. Also, we generate embedding vectors 
of length~100. The value of~100 was chosen after experimenting with different values of length.

\begin{table}[!htb]
\centering
\caption{Summary of the LSTM models}\label{tab: summary_lstm_models}
\adjustbox{scale=0.85}{
\begin{tabular}{c|cc}\midrule\midrule
Criteria & Word2Vec-LSTM & BERT-LSTM\\ \midrule
Trainable Parameters & 16,359 & 96,519 \\
Non-trainable Parameters & 42,500 & 121,344 \\
Total Parameters & 58,859 & 217,863 \\ \midrule\midrule
\end{tabular}
}
\end{table}

Based on the experiments summarized in Table~\ref{tab: k_lstm_word2vec},
we found that length~$k=2500$
gave us the best results for both LSTM models. Hence, we use this value of~$k$ for both the Word2Vec-LSTM and
BERT-LSTM results reported in this paper. It can be observed that the length of the observation sequence taken into consideration for LSTM models (i.e.,$k$) is significantly greater than other models such as HMM-RF (i.e., $L$). This is because of the fact that LSTM models excel at understanding context over a large sequence of inputs, whereas in HMM-RF, we have already captured these long-term dependencies while training the HMM.

\begin{table}[!htb]
\centering
\caption{Accuracy of LSTM models as a function of sequence length}\label{tab: k_lstm_word2vec}
\adjustbox{scale=0.85}{
\begin{tabular}{c|cc}\midrule\midrule
$k$ & Word2Vec-LSTM & BERT-LSTM\\ \midrule
1000 & 0.9708 & 0.9150 \\
2500 & 0.9714 & 0.9181 \\
5000 & 0.9696 & 0.9119 \\ \midrule\midrule
\end{tabular}
}
\end{table}

\begin{figure}[!htb]
\centering
\includegraphics[width=150mm, height = 70mm]{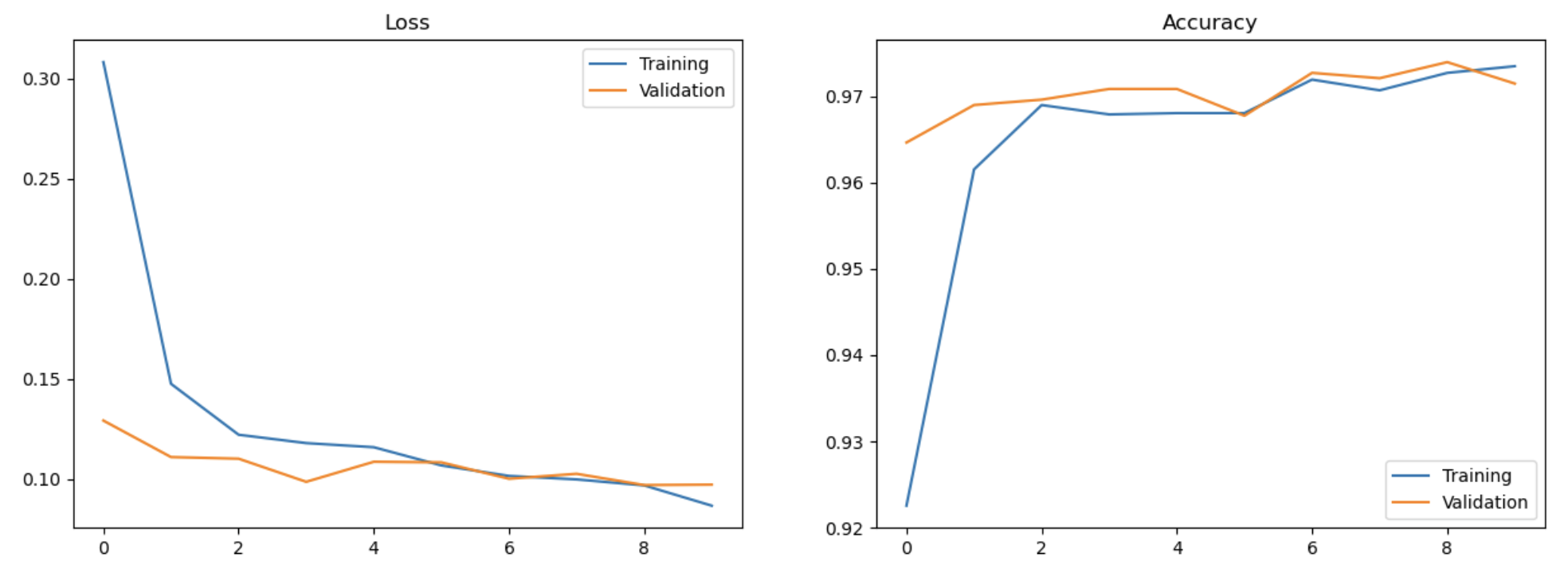}
\caption{Accuracy and loss graphs for Word2Vec-LSTM model}\label{fig:Graph_Word2Vec}
\end{figure}

\begin{figure}[!htb]
\centering
\includegraphics[width=150mm, height = 70mm]{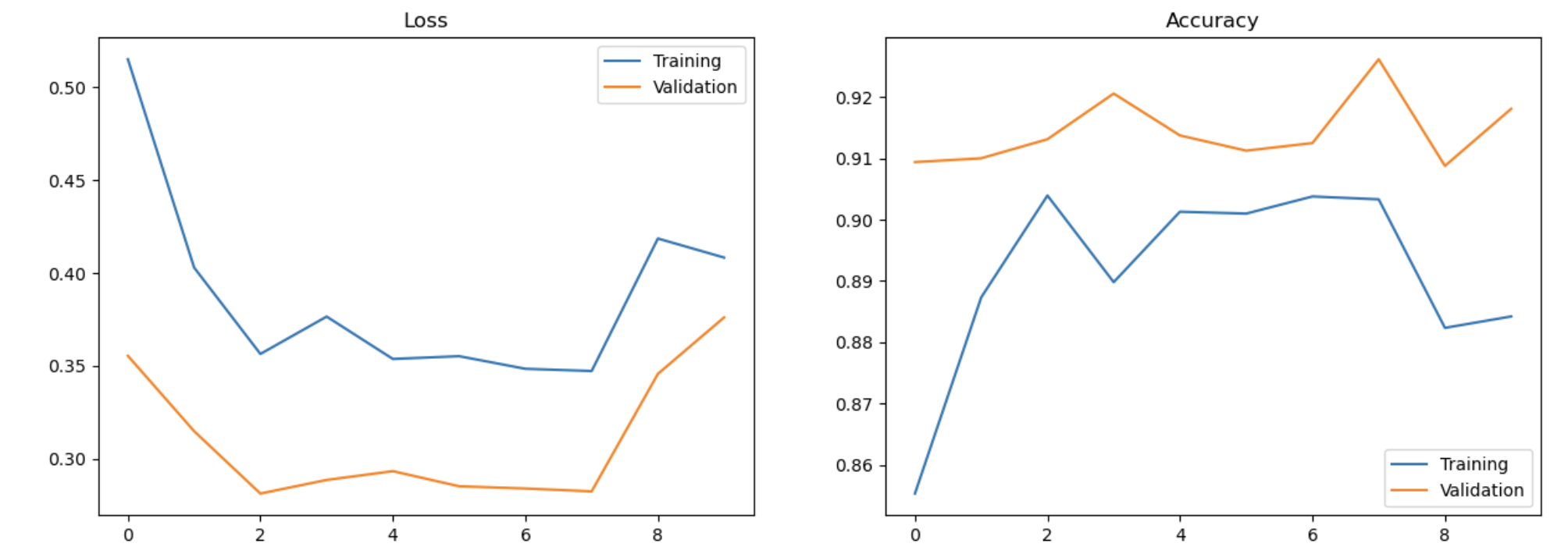}
\caption{Accuracy and loss graphs for BERT-LSTM model}\label{fig:Graph_BERT}
\end{figure}

Figure~\ref{fig:Graph_Word2Vec} shows accuracy and loss graphs for the Word2Vec-LSTM 
model, while Figure~\ref{fig:Graph_BERT} shows the analogous graphs for the BERT-LSTM model.
These graphs show that Word2Vec-LSTM model is very well-behaved, with no indication
of overfitting. On the other hand, the BERT-LSTM model is not as well-behaved.

\end{document}

%% file: preamble.tex
\usepackage{amsmath,amsthm, amsfonts, amssymb, amsxtra, amsopn}
\usepackage{pgfplots}
\usetikzlibrary{calc}
\pgfplotsset{compat=1.13}
\usepackage{pgfplotstable}
\usepackage{graphicx,grffile}
\usepackage{multirow}
\usepackage{float}
\usepackage{booktabs}
\usepackage{listings}
\usepackage{cmap}
\usepackage{colortbl}
\usepackage{adjustbox}
\usepackage{epsfig}
\usepackage{bold-extra}

\usepackage[tableposition=top,font=small,skip=5pt]{caption}
\usepackage{subcaption}
\usepackage{makecell} 

\def\cO{{\cal O}}

\PassOptionsToPackage{hyphens}{url}

\usepackage{hyperref}
\hypersetup{colorlinks=true,linkcolor=black,citecolor=black,urlcolor=blue,filecolor=black}
\hypersetup{pdfpagemode=UseNone,pdfstartview=}

\usepackage{enumitem}
\setlist[itemize]{noitemsep, topsep=0pt}

\advance\oddsidemargin by -0.35in
\advance\textwidth by 0.7in

\advance\topmargin by -0.4in
\advance\textheight by 0.8in

\long\def\symbolfootnotetext[#1]#2{\begingroup%
\def\thefootnote{\fnsymbol{footnote}}\footnotetext[#1]{#2}\endgroup}

\let\vv=\v
\def\v{{\tt v}}

\def\cO{{\cal O}}

\def\zz{\phantom{0}}

%% file: figures/hmm.tex
    \begin{tikzpicture}[scale=1.0]
    
    \draw[thick,color=blue] (0,0) rectangle (1,1);
    \draw[thick,color=blue] (2.5,0) rectangle (3.5,1);
    \draw[thick,color=blue] (5,0) rectangle (6,1);
    \draw[thick,color=blue] (10,0) rectangle (11,1);

    \draw[thick,color=green] (0.5,4.5) circle (0.575);
    \draw[thick,color=green] (3,4.5) circle (0.575);
    \draw[thick,color=green] (5.5,4.5) circle (0.575);
    \draw[thick,color=green] (10.5,4.5) circle (0.575);
    
    \node at (0.5,0.5){$\cO_0$};
    \node at (3,0.5){$\cO_1$};
    \node at (5.5,0.5){$\cO_2$};
    \node at (8,0.5){$\cdots$};
    \node at (10.5,0.5){$\cO_{T-1}$};

    \node at (0.5,4.5){$X_0$};
    \node at (3,4.5){$X_1$};
    \node at (5.5,4.5){$X_2$};
    \node at (8,4.5){$\cdots$};
    \node at (10.5,4.5){$X_{T-1}$};
       
    \node at (1.7,4.8){$A$};
    \node at (4.2,4.8){$A$};
    \node at (6.7,4.8){$A$};
    \node at (9.2,4.8){$A$};
    
    \node at (0.2,2.1){$B$};
    \node at (2.7,2.1){$B$};
    \node at (5.2,2.1){$B$};
    \node at (10.2,2.1){$B$};
    
     \draw[thick,color=black,->] (1.075,4.5) -- (2.425,4.5);
     \draw[thick,color=black,->] (3.575,4.5) -- (4.925,4.5);
     \draw[thick,color=black,->] (6.075,4.5) -- (7.425,4.5);
     \draw[thick,color=black,->] (8.575,4.5) -- (9.925,4.5);

     \draw[thick,color=black,->] (0.5,3.925) -- (0.5,1);
     \draw[thick,color=black,->] (3.0,3.925) -- (3.0,1);
     \draw[thick,color=black,->] (5.5,3.925) -- (5.5,1);
     \draw[thick,color=black,->] (10.5,3.925) -- (10.5,1);

    \draw[thick,dashed,color=red] (-0.3,3) -- (11.2,3);
   
    \end{tikzpicture}

%% file: figures/tree.tex
    \begin{tikzpicture}[scale=0.95,every node/.style={scale=0.9}]
    \draw[thick,color=black] (0,0) -- (2.5,2.5);
    \draw[thick,color=black] (0,0) -- (2.5,-2.5);
    \draw[thick,color=black] (4.0,2.5) -- (6.5,3.75);
    \draw[thick,color=black] (4.0,2.5) -- (6.5,1.25);
    \draw[thick,color=black] (4.0,-2.5) -- (6.5,-1.25);
    \draw[thick,color=black] (4.0,-2.5) -- (6.5,-3.75);
    \node at (-0.75,0){file size};
    \node at (3.25,2.5){entropy};
    \node at (3.25,-2.5){entropy};
    \node[color=blue] at (7.2,3.75){benign};
    \node[color=red] at (7.35,-1.2){\underline{malware}};
    \node[color=blue] at (7.2,1.25){benign};
    \node[color=blue] at (7.2,-3.75){benign};

   \node[rotate=45] at (1.1,1.4){large};
   \node[rotate=-45] at (1.415,-1.085){small};
   \node[rotate=28] at (5.2,3.325){high};
   \node[rotate=28] at (5.2,-1.675){high};
   \node[rotate=-28] at (5.45,-2.96){low};
   \node[rotate=-28] at (5.45,2.04){low};

    

    \end{tikzpicture}

%% file: figures/dist.tex
\begin{tikzpicture}[scale=1.0, every node/.style={scale=1.0}]
\pgfkeys{/pgf/number format/.cd,1000 sep={}}
\begin{axis}[
        width  = 0.7*\textwidth,
        height = 7.5cm,
        ymin=0,ymax=5000,
        ytick={0,1000,2000,3000,4000,5000},
        major x tick style = transparent,
        ybar=5*\pgflinewidth,
        bar width=25.0pt,
        ylabel = {Samples},
        symbolic x coords={ZeroAccess,Winwebsec,SecurityShield,Zbot,Cridex,SmartHDD,Harebot},
        xticklabels={ZeroAccess,Winwebsec,SecurityShield,Zbot,Cridex,SmartHDD,Harebot},
        label style={scale=0.90},
	y tick label style={
		scale=0.80,
    		/pgf/number format/.cd,
   		fixed,
   		fixed zerofill,
    		precision=0},
        xtick = data,
        x tick label style={
        		rotate=60,
		scale=0.80,
		anchor=north east,
		inner sep=0mm
		},
        nodes near coords,
        every node near coord/.append style={
								   scale=0.80,
								   /pgf/number format/.cd,
								   fixed,
								   fixed zerofill,
								   precision=0},
        enlarge x limits=0.12,
        legend cell align=left,
        legend style={
                at={(0.91,0.02)},
                anchor=south,
                column sep=1ex
        },
]
\addplot [fill=blue,opacity=1.00]
coordinates {
(ZeroAccess, 1305)
(Winwebsec, 4360)
(SecurityShield, 58)
(Zbot, 2136)
(Cridex, 74)
(SmartHDD, 68)
(Harebot, 53)
};
\end{axis}
\end{tikzpicture}

%% file: figures/line.tex
\begin{tikzpicture}[scale=0.65, every node/.style={scale=0.95}]
\begin{axis}[
		   width=0.7\textwidth,
		   height=0.55\textwidth,
	 	   x tick label style={
		   },
	 	   y tick label style={
    		 	/pgf/number format/.cd,
   			fixed,
   			fixed zerofill,
    			precision=2
		    },
	        	   ymin=0.86,ymax=1.0,
        		   ytick={0.86,0.88,0.90,0.92,0.94,0.96,0.98,1.00},
        		   major x tick style = transparent,
                    legend pos=south east,
                    xlabel={n{\textunderscore}estimators},
                    ylabel={Training accuracy},
                    xtick = data,
                    symbolic x coords={1, 10, 100, 150, 200},
                    xticklabels={1, 10, 100, 150, 200}] 
\addplot[color=blue,ultra thick,mark=*,mark size=2.0] coordinates {
(1, 0.9491459627329193)
(10, 0.961516563146998)
(100, 0.9628364389233955)
(150, 0.9625517598343686)
(200,0.9623447204968945)
};
\addplot[color=red,ultra thick,mark=*,mark size=2.0] coordinates {
(1,0.952155361486621)
(10,0.966247634054788)
(100, 0.9678016557487786)
(150, 0.9682680978948475)
(200, 0.9681643970913022)
};
\addplot[color=yellow,ultra thick,mark=*,mark size=2.0] coordinates {
(1, 0.943253425912835)
(10, 0.962454226839919)
(100,0.9644239379076082)
(150, 0.9642683162123286)
(200, 0.9641127952169942)
};
\addplot[color=black,ultra thick,mark=*,mark size=2.0] coordinates {
(1, 0.9310430278272745)
(10, 0.9581385989793568)
(100, 0.9614607399865046)
(150, 0.9610452097247979)
(200, 0.9613047747190807)
};
\legend{$L=25$, $L=50$, $L=100$, $L=200$}
\end{axis}
\end{tikzpicture}

%% file: figures/criterion.tex
\begin{tikzpicture}[scale=0.65, every node/.style={scale=0.95}]
\pgfkeys{/pgf/number format/.cd,1000 sep={}}
\begin{axis}[
        width  = 0.7*\textwidth,
        height = 7.5cm,
        ymin=0.85,ymax=1.0,
        ytick={0.86,0.88,0.90,0.92,0.94,0.96,0.98,1.00},
        major x tick style = transparent,
        ybar=5*\pgflinewidth,
        bar width=24.0pt,
        xlabel = {Criterion},
        ylabel = {Training accuracy},
        symbolic x coords={Gini, Entropy},
        xticklabels={Gini, Entropy},
	y tick label style={
    		/pgf/number format/.cd,
   		fixed,
   		fixed zerofill,
    		precision=2},
        xtick = data,
        x tick label style={
		font=\small,
		},
        enlarge x limits=0.5,
        legend cell align=left,
        legend style={
                at={(0.86,0.02)},
                anchor=south,
                column sep=1ex
        },
]
\addplot [fill=blue,opacity=1.00]
coordinates {
(Gini, 0.9594099378881987)
(Entropy, 0.9599482401656315)
};
\addlegendentry{$L=25$}
\addplot [fill=red,opacity=1.00]
coordinates {
(Gini, 0.9644394094471855)
(Entropy, 0.9646154490633493)
};
\addlegendentry{$L=50$}
\addplot [fill=yellow,opacity=1.00]
coordinates {
(Gini, 0.9600029726623817)
(Entropy, 0.9594021081734923)
};
\addlegendentry{$L=100$}
\addplot [fill=black,opacity=1.00]
coordinates {
(Gini, 0.9552160684630958)
(Entropy, 0.9539808720317102)
};
\addlegendentry{$L=200$}
\end{axis}
\end{tikzpicture}

%% file: figures/max.tex
\begin{tikzpicture}[scale=0.65, every node/.style={scale=0.95}]
\pgfkeys{/pgf/number format/.cd,1000 sep={}}
\begin{axis}[
        width  = 0.7*\textwidth,
        height = 7.5cm,
        ymin=0.85,ymax=1.0,
        ytick={0.86,0.88,0.90,0.92,0.94,0.96,0.98,1.00},
        major x tick style = transparent,
        ybar=5*\pgflinewidth,
        bar width=15.0pt,
        xlabel = {Max feature},
        ylabel = {Training accuracy},
        symbolic x coords={Sqrt, Log2, None},
        xticklabels={Sqrt, Log2, None},
	y tick label style={
    		/pgf/number format/.cd,
   		fixed,
   		fixed zerofill,
    		precision=2},
        xtick = data,
        x tick label style={
		font=\small,
		},
        enlarge x limits=0.25,
        legend cell align=left,
        legend style={
                at={(0.86,0.02)},
                anchor=south,
                column sep=1ex
        },
]
\addplot [fill=blue,opacity=1.00]
coordinates {
(Sqrt, 0.9600621118012421)
(Log2, 0.9597826086956524)
(None, 0.9591925465838511)
};
\addlegendentry{$L=25$}
\addplot [fill=red,opacity=1.00]
coordinates {
(Sqrt, 0.9652319945477024)
(Log2, 0.96283844546675)
(None, 0.96551184775135)
};
\addlegendentry{$L=50$}
\addplot [fill=yellow,opacity=1.00]
coordinates {
(Sqrt, 0.9604488357676539)
(Log2, 0.9572772708542537)
(None, 0.9613815146319036)
};
\addlegendentry{$L=100$}
\addplot [fill=black,opacity=1.00]
coordinates {
(Sqrt, 0.9558233269088573)
(Log2, 0.9495952216444234)
(None, 0.958376862188928)
};
\addlegendentry{$L=200$}
\end{axis}
\end{tikzpicture}

%% file: figures/conf_HMM-RF.tex
\begin{tikzpicture}[scale=0.525]
    \begin{axis}[
        width=10cm,
        height=10cm,
	colormap={bluewhite}{color=(white) rgb255=(100,149,237)},
        xticklabels={ZeroAccess,Winwebsec,SecurityShield,Zbot,Cridex,SmartHDD,Harebot},
        xtick={0,...,6},
        xtick style={draw=none},
	xticklabel style={anchor=east,rotate=45,yshift=-5pt,font=\large},
        yticklabels={ZeroAccess,Winwebsec,SecurityShield,Zbot,Cridex,SmartHDD,Harebot},
        ytick={0,...,6},
        ytick style={draw=none},
        enlargelimits=false,
        yticklabel style={font=\large},
        colorbar,
        colorbar style={
            ytick={0,40,80,120,160,200},
            yticklabels={0,40,80,120,160,200},
            yticklabel={\pgfmathprintnumber\tick},
            yticklabel style={
            		scale=1.33,
            		/pgf/number format/fixed,
			/pgf/number format/precision=0}
        },
        point meta min=0,
        point meta max=200,
        nodes near coords={\pgfmathprintnumber\pgfplotspointmeta},
        nodes near coords black white/.style={
            small value/.style={
                yshift=-7pt,
                text=black,
                /pgf/number format/fixed,
                /pgf/number format/precision=0,
                /pgf/number format/zerofill=true,
                scale=1.2,
            },
            large value/.style={
                yshift=-7pt,
                text=white,
                /pgf/number format/fixed,
                /pgf/number format/precision=0,
                /pgf/number format/zerofill=true,
                scale=1.2,
            },
            every node near coord/.style={
                check for zero/.code={
                    \pgfmathfloatifflags{\pgfplotspointmeta}{0}{
                        \pgfkeys{/tikz/coordinate}
                    }{
                        \begingroup
                        \pgfkeys{/pgf/fpu}
                        \pgfmathparse{\pgfplotspointmeta<#1}
                        \global\let\result=\pgfmathresult
                        \endgroup
                        %
                        %
                        \pgfmathfloatcreate{1}{1.0}{0}
                        \let\ONE=\pgfmathresult
                        \ifx\result\ONE
                            \pgfkeysalso{/pgfplots/small value}
                        \else
                            \pgfkeysalso{/pgfplots/large value}
                        \fi
                    }
                },
                check for zero,
            },
        },
        nodes near coords black white=100,
    ]
        \addplot[
            matrix plot,
            mesh/cols=7,
            point meta=explicit,draw=gray
        ] table [meta=C] {
            x y C
0 0 259
1 0 2
2 0 0
3 0 0
4 0 0
5 0 0
6 0 0
0 1 0
1 1 870
2 1 0
3 1 0
4 1 0
5 1 0
6 1 0
0 2 0
1 2 2
2 2 10
3 2 0
4 2 0
5 2 0
6 2 0
0 3 5
1 3 13
2 3 0
3 3 409
4 3 0
5 3 0
6 3 0
0 4 1
1 4 10
2 4 0
3 4 1
4 4 3
5 4 0
6 4 0
0 5 0
1 5 0
2 5 0
3 5 0
4 5 0
5 5 14
6 5 0
0 6 0
1 6 4
2 6 0
3 6 1
4 6 0
5 6 0
6 6 5
         };
    \end{axis}
\end{tikzpicture}

%% file: figures/conf_relative_HMM-RF.tex
\begin{tikzpicture}[scale=0.525]
    \begin{axis}[
        width=10cm,
        height=10cm,
	colormap={bluewhite}{color=(white) rgb255=(100,149,237)},
        xticklabels={ZeroAccess,Winwebsec,SecurityShield,Zbot,Cridex,SmartHDD,Harebot},
        xtick={0,...,6},
        xtick style={draw=none},
	xticklabel style={anchor=east,rotate=45,yshift=-5pt,font=\large},
        yticklabels={ZeroAccess,Winwebsec,SecurityShield,Zbot,Cridex,SmartHDD,Harebot},
        ytick={0,...,6},
        ytick style={draw=none},
        enlargelimits=false,
        yticklabel style={font=\large},
        colorbar,
        colorbar style={
            ytick={0,0.2,0.4,0.6,0.8,1.0},
            yticklabels={0,0.2,0.4,0.6,0.8,1.0},
            yticklabel={\pgfmathprintnumber\tick},
            yticklabel style={
            		scale=1.33,
            		/pgf/number format/fixed,
			/pgf/number format/precision=2}
        },
        point meta min=0.0,
        point meta max=1.0,
        nodes near coords={\pgfmathprintnumber\pgfplotspointmeta},
        nodes near coords black white/.style={
            small value/.style={
                yshift=-7pt,
                text=black,
                /pgf/number format/fixed,
                /pgf/number format/precision=3,
                /pgf/number format/zerofill=true,
                scale=1.0,
            },
            large value/.style={
                yshift=-7pt,
                text=white,
                /pgf/number format/fixed,
                /pgf/number format/precision=3,
                /pgf/number format/zerofill=true,
                scale=1.0,
            },
            every node near coord/.style={
                check for zero/.code={
                    \pgfmathfloatifflags{\pgfplotspointmeta}{0}{
                        \pgfkeys{/tikz/coordinate}
                    }{
                        \begingroup
                        \pgfkeys{/pgf/fpu}
                        \pgfmathparse{\pgfplotspointmeta<#1}
                        \global\let\result=\pgfmathresult
                        \endgroup
                        %
                        %
                        \pgfmathfloatcreate{1}{1.0}{0}
                        \let\ONE=\pgfmathresult
                        \ifx\result\ONE
                            \pgfkeysalso{/pgfplots/small value}
                        \else
                            \pgfkeysalso{/pgfplots/large value}
                        \fi
                    }
                },
                check for zero,
            },
        },
        nodes near coords black white=0.5,
    ]
        \addplot[
            matrix plot,
            mesh/cols=7,
            point meta=explicit,draw=gray
        ] table [meta=C] {
            x y C
0 0 0.9923
1 0 0.0077
2 0 0
3 0 0
4 0 0
5 0 0
6 0 0
0 1 0
1 1 1.0000
2 1 0
3 1 0
4 1 0
5 1 0
6 1 0
0 2 0
1 2 0.1667
2 2 0.8333
3 2 0
4 2 0
5 2 0
6 2 0
0 3 0.0118
1 3 0.0304
2 3 0
3 3 0.9578
4 3 0
5 3 0
6 3 0
0 4 0.0667
1 4 0.6666
2 4 0
3 4 0.0667
4 4 0.2000
5 4 0
6 4 0
0 5 0
1 5 0
2 5 0
3 5 0
4 5 0
5 5 1.0000
6 5 0
0 6 0
1 6 0.4000
2 6 0
3 6 0.1000
4 6 0
5 6 0
6 6 0.5000
         };
    \end{axis}
\end{tikzpicture}